\documentclass[conference]{IEEEtran}
\IEEEoverridecommandlockouts
\usepackage{amsmath,amssymb,amsfonts}
\usepackage{algorithmic, algorithm}
\usepackage{cite, amsfonts, amssymb, amsthm, bm, bbm, graphicx, relsize, multirow, booktabs, tikz,soul}
\usepackage{subcaption}
\usepackage{tabulary}
\usepackage{mathbbol}
\usepackage{textcomp}
\usepackage{xcolor}
\def\BibTeX{{\rm B\kern-.05em{\sc i\kern-.025em b}\kern-.08em T\kern-.1667em\lower.7ex\hbox{E}\kern-.125emX}}

\newcommand{\ds}{\displaystyle}

\newcommand{\bH}{\mathbf{H}}
\newcommand{\ba}{\mathbf{a}}
\newcommand{\bx}{\mathbf{x}}

\newcommand{\bW}{\mathbf{W}}
\newcommand{\bu}{\mathbf{u}}
\newcommand{\bb}{\mathbf{b}}
\newcommand{\br}{\mathbf{r}}
\newcommand{\bz}{\mathbf{z}}
\newcommand{\bv}{\mathbf{v}}

\def\BibTeX{{\rm B\kern-.05em{\sc i\kern-.025em b}\kern-.08em
    T\kern-.1667em\lower.7ex\hbox{E}\kern-.125emX}}
\begin{document}

\title{Beam Alignment in mmWave User-Centric Cell-Free Massive MIMO Systems
}

\author{\IEEEauthorblockN{Stefano Buzzi, Carmen D'Andrea}
	\IEEEauthorblockA{University of Cassino and Southern Latium\\
		Dept. of Electrical and Information Eng.\\
		Cassino, Italy}
	\and
	\IEEEauthorblockN{Maria Fresia, Xiaofeng Wu}
	\IEEEauthorblockA{Huawei Technologies Duesseldorf GmbH\\
		Wireless Terminal Chipset Technology Lab\\
		Munich, Germany}}

\maketitle

\begin{abstract}
The problem of beam alignment (BA) in a cell-free massive multiple-input multiple-output (CF-mMIMO) system operating at millimeter wave (mmWaves) carrier frequencies is considered in this paper. Two estimation algorithms are proposed, in association with a protocol that permits simultaneous estimation, on a shared set of frequencies, for each user equipment (UE), of the direction of arrival and departure of the radio waves associated to the strongest propagation paths from each of the surrounding access points (APs), so that UE-AP association can take place.
The proposed procedure relies on the existence of a reliable control channel at sub-6 GHz frequency, so as to enable exchange of estimated values between the UEs and the network, and assumes that APs can be identifies based on the prior knowledge of the orthogonal channels and transmit beamforming codebook. 
A strategy for assigning codebook entries to the several APs is also proposed, with the aim of minimizing the mutual interference between APs that are assigned the same entry. 
Numerical results show the effectiveness of the proposed detection strategy, thus enabling one shot fast BA for CF-mMIMO systems. 
\end{abstract}

\begin{IEEEkeywords}
cell-free massive MIMO, user-centric, beam alignment, millimeter wave
\end{IEEEkeywords}

\section{Introduction}
In recent years, mmWave carrier frequencies have been widely considered for adoption in wireless cellular communications due to the availability of large bandwidths.  In order to overcome the large path loss that is typical of such frequencies, multiple antennas and narrow beams must be used so as to concentrate radiated energy along  spatial directions associated with the strongest propagation paths \cite{Heath_SP_mmwave_2016}. The problem of finding those beamforming directions is usually called ``\textit{beam alignment}'' (BA). Performing BA is a pre-requisite in order to have reliable data transmission, and, as such, has been widely investigated in recent years -- see for instance \cite{Caire_scalable_robust_BA_TCOM2018,hassanieh2018fast,li2019fast,song2015adaptive,maschietti2017robust}. Notably, the paper \cite{Caire_scalable_robust_BA_TCOM2018} proposes a BA algorithm wherein a (single) base station transmits a sequence of pseudo-random multi-finger beam patters and the UEs, in receive mode, use the non-negative constrained least-squares algorithm  to estimate the directions of arrival and departure of the strongest beam. The majority of existing solutions, however, typically refer to a scenario involving a single AP and a  single UE, and BA is performed usually using a protocol that realizes several interactions between the AP and the UE. In future beyond-5G and 6G wireless networks, however, CF-mMIMO network deployments \cite{Ngo_CellFree2017, Buzzi_WCL_cellfree}, wherein multiple low-complexity APs jointly serve several UEs in the same frequency band, will be used very frequently. It is thus of primary interest develop BA algorithms able to operate in the presence of multiple APs and multiple UEs sharing the same communication bandwidth. 

The aim of this paper is to address the BA problem in the relevant multi-AP multi-UE scenario. Departing from the procedure in \cite{Caire_scalable_robust_BA_TCOM2018}, we develop a protocol, where, thanks to the aid of a sub-6 GHz uplink/downlink control channel, BA is performed \textit{simultaneously} for all the UEs in the system, using the same set of carrier frequencies. Two different estimation algorithms, to be run at the UE, are proposed, that permit to estimate, for each surrounding AP, the direction of arrival and of departure of the strongest beam, and its associated strength.  In order to be able to distinguish the several APs, these must transmit on disjoint set of orthogonal channels. Since, in general, the number of APs is much greater than the number of available sets of orthogonal channels, we also propose an algorithm to assign these sets to the APs, with the objective of minimizing the mutual interference among APs using the same set of orthogonal carriers. The performance of the proposed protocol and estimation algorithms is numerically assessed, in terms of probability of correct detection of the direction of arrival and departure of the strongest beams, and numerical results will show the effectiveness of the proposed approaches.

\section{System description}
\begin{figure}
	\begin{center}
		\includegraphics[scale=0.3]{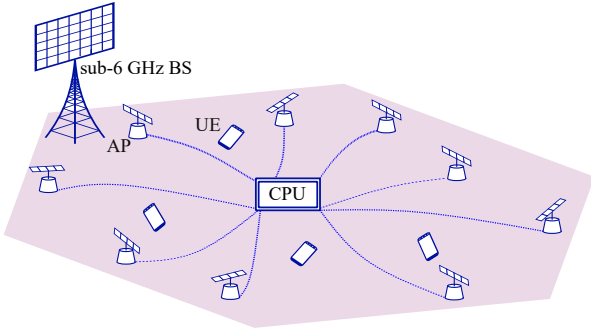}
	\end{center}
	\caption{Considered scenario.}
	\label{Fig:scenario}
\end{figure}

Consider a cell-free massive MIMO system where $M$ APs are distributed in a given area in order to simultaneously serve $K$ UEs on a shared channel. We focus on the BA procedure and, for the sake of simplicity, consider a bi-dimensional layout\footnote{Extension to 3D layouts can be straightforwardly done.}. It is assumed that the adopted modulation format is the orthogonal frequency division multiplexing (OFDM); we denote by $t_0$ the duration of an OFDM symbol, by $B$ the overall available bandwidth, and by $\Delta f$ the subcarrier spacing for the OFDM signal. This implies that the number of subcarriers is $N_C=B/\Delta f$. The duration of the OFDM symbol is taken equal to $1/\Delta f + \tau_{CP}$, with $\tau_{CP}$ the length of the cyclic prefix, chosen not smaller than the largest multipath delay spread in the system. The BA phase will span $T$ \textit{beacon slots}, each made of $S$ OFDM symbols. 
We denote as $N_{\rm UE}$, $N_{\rm AP}$, $n_{\rm UE}< N_{\rm UE}$ and $n_{\rm AP}< N_{\rm AP}$ the numbers of antennas  and RF chains at the generic UE and AP, respectively.
Both the APs and the UEs are equipped with uniform linear arrays (ULAs) with random orientations, and the steering angles are assumed to take value in the range $[-\pi/2,\pi/2]$. A representation of the considered scenario is reported in Fig. \ref{Fig:scenario}. 
\subsection*{Channel model}
The downlink channel between the $m$-th AP and the $k$-th UE in the $s$-th beacon slot is expressed through a matrix-valued linear time invariant (LTI) channel whose impulse response is the following matrix-valued $(N_{\rm UE} \times N_{\rm AP})$-dimensional function:
\begin{equation}
\bH_{k,m}^{(s)}(\tau)\!\!=\!\!\ds \sum_{\ell=0}^{L_{k,m}} \!\!\alpha_{k,m}^{(s)}(\ell) 
\ba_{\rm UE}(\varphi_{k,m,\ell}) \ba_{\rm AP}^H(\theta_{k,m,\ell}) \delta(\tau-\tau_{k,m,\ell}) \; ,
\label{eq:channelmodel_km} 
\end{equation} 
where $L_{k,m}$ denotes the number of paths that contribute to the channel between the $k$-th UE and the $m$-th AP, $\alpha_{k,m}^{(s)}(\ell) \sim \mathcal{CN}(0, \gamma_{k,m}(\ell))$ is the complex gain associated to the $\ell$-th path in the $s$-th beacon slot, $\gamma_{k,m}(\ell)$, $\varphi_{k,m,\ell}$, $\theta_{k,m,\ell}$ and $\tau_{k,m,\ell}$ are the variance of the complex gain (including path loss and shadow fading), the angle of arrival (AoA), angle of departure (AoD) and the propagation delay associated to the $\ell$-th path between the $k$-th UE and the $m$-th AP and $\ba_{\rm AP}(\cdot)$ and $\ba_{\rm UE}(\cdot)$ are the ULA array responses at the AP and at the UE, respectively. Notice that $L_{k,m}$ depends on the geometry of the system, usually we have that $L_{k,m} << \min\{N_{\rm AP}, N_{\rm UE}\}$. Additionally, while the complex gains $\alpha_{k,m}^{(s)}(\ell)$ depend on the beacon slot index $s$, this does not happen for the other parameters which typically vary over much larger timescales than the complex gains associated with propagation paths.

\section{Beam Alignment procedure description} \label{BA_procedure}
We now focus on the description of the proposed BA algorithms; we start illustrating preliminary steps and working assumptions that are needed in order to set up the system and realize the BA procedure.

Before the BA procedure starts, the \textit{data pattern} are to be defined and assigned to the APs. Assuming the use of OFDM, we define as data pattern a set of subcarriers and of beamforming vectors. 
Indeed, since each AP is equipped with $n_{\rm AP}$ RF chains, i.e., it can simultaneously transmit $n_{\rm AP}$ data streams using different beamforming vectors. In order to permit data stream separation at the UEs without having knowledge of the used beamformers, it is needed that the transmitted data streams  are orthogonal \textit{before} beamforming. This can be achieved by letting the data streams use non-overlapping subcarriers. This poses a problem about the availability of a sufficient number of orthogonal (i.e. non-overlapping) data patterns. Ideally, data streams transmitted by different APs should be orthogonal in order to permit to the UEs to separate the measurements coming from different APs. In practice however, due to the unavailability of a sufficient number of orthogonal data pattern, the same set of these resources is to be reused across the network.
With regard to the subcarriers, denoting by $Q$ the number of subcarriers assigned for each AP RF chain, and by $N_C$ the total number of subcarriers, we can accordingly define $D= \lfloor \lfloor N_C/Q\rfloor / n_{\rm AP} \rfloor$ disjoint sets of carrier frequencies; the corresponding $D$ data patterns are denoted by  ${\cal D}_1, \ldots, {\cal D}_D$. More precisely, the generic set ${\cal D}_d$ will specify, for each of the $n_{\rm AP}$ transmit RF chains, the $Q$ subcarriers \textit{and} the beamformers to be used in each slot. Otherwise stated, letting $\mathcal{L}_{d,s,i}$ denote the set of $Q$ subcarriers to be used by the APs that are assigned the $d$-th data pattern in the $s$-th beacon slot and on the $i$-th RF chain, the data pattern ${\cal D}_d$ is formally described as
\begin{equation}
	\begin{array}{llll}
		{\cal D}_d=& \left\{ \left\{ \mathcal{L}_{d,s,i}, \, s=1, \ldots, T , \; i=1, \ldots, n_{\rm AP} \right\} ,\, \; \right.
		\\ & \left. \; \left\{ \mathcal{B}_{d,s,i}, \, s=1, \ldots, T , \; i=1, \ldots, n_{\rm AP} \right\}
		\right\} \; ,
	\end{array}
	\label{eq:Dd}
\end{equation}
where $\mathcal{B}_{d,s,i}$  denotes the beamformer used by the APs that have been assigned the $d$-th data pattern in the $s$-th beacon slot and on the $i$-th RF chain. These beamformers will be specified later.
We emphasize that the subcarrier indexes in the sets $\mathcal{L}_{d,s,i}$ are disjoint, i.e., in each beacon slot, each subcarrier can be in only one of the sets $\mathcal{L}_{d,s,i}$, so that orthogonality can be preserved. The $D$ data pattern ${\cal D}_1, \ldots, {\cal D}_D$, as previously discussed, are to be shared among the $M>D$ APs. This poses the problem of how to assign data patterns to the APs, which is considered next.

\subsection{Location-based data pattern assignment}
A location-based (LB) data pattern assignment is here proposed to allocate the data patterns to the APs in order to reduce the \textit{BA contamination} in the system. The proposed method uses the k-means clustering method, i.e., an iterative algorithm that serves to partition APs into $D$ APs groups using the same data pattern. The algorithm, accepting as input the APs positions and the number $D$ of data patterns,  is summarized, for the sake of brevity, in Algorithm \ref{LB_resource_set}.

\subsection{Beam alignment procedure phases}
We assume that there is a general frame synchronization information available in the system that can be ensured by exploiting the fronthaul connection between the APs and the CPU, and by using a control-plane connection with the UEs at a sub-6 GHz carrier frequency. The BA procedure is made of three phases: 
\begin{itemize}
	\item[a)] All the APs transmit simultaneously proper signals using proper beamforming schemes and the UEs gather information and estimate the AoAs and AoDs corresponding to the strongest received paths from the $D$ surrounding APs; 
	\item[b)] Using the sub-6 GHz uplink control channel, each UE communicates to the network its position and, for each of the $D$ data patterns, the AoA and AoD of the strongest beam and a strength indicator;
	\item[c)] Based on the information gathered from all the UEs, the network makes user-centric AP-UE association and communicates AP-UE associations to the APs and to the UEs (via the sub-6 GHz control channel);  
	\item[d)] the communication between the BSs and the UEs can start using the beams that have been determined in the previous two phases.
\end{itemize} 

\begin{algorithm}
	
	\caption{Location-based data patterns assignment algorithm}
	
	\begin{algorithmic}[1]
		
		\label{LB_resource_set}
		\STATE  Allocate randomly  $\lfloor M/ D\rfloor$ {centroids}.
		\REPEAT 
		
		\STATE Assign each AP to the nearest centroid with the constraint that no more than $D$ APs are associated to the same centroid.
		
		\STATE Compute the new positions of the centroids averaging over the positions of the APs belonging to the same cluster.
		
		\UNTIL convergence of the positions of the centroids or maximum number of iterations reached. 
		
		\STATE Assign the first data pattern to the AP in each cluster that has the largest latitude (i.e. the most northern one); assign the second data pattern to each AP in the cluster with the second largest latitude. Continue until all the $D$ data patterns have been assigned.
		
	\end{algorithmic}
	
\end{algorithm}

\section{Transmitted signal model and signal processing at the UE}

\subsection{Signal model}
Let us focus on the signal transmitted in the $s$-th beacon slot, i.e. for $t \in [sTt_0, (s+1)Tt_0]$. The baseband equivalent of the signal transmitted in the $s$-th beacon slot by the $m$-th AP can be expressed through the following $N_{\rm AP}$-dimensional vector-valued waveform:
\begin{equation}
	\bx_{m,s}(t)=\ds \sum_{i=1}^{n_{\rm AP}}x_{m,s,i}(t) \bu_{m,s,i} 	\; ,
\end{equation}
where $x_{m,s,i}(t)$ is the signal corresponding to the $i$-th data stream from the $m$-th AP in the $s$-th beacon interval; the  $N_{\rm AP}$-dimensional vector $\bu_{m,s,i}$  is the corresponding transmit beamformer\footnote{Notice that we are here implicitly assuming that the transmit beamformer is kept constant over an entire beacon slot, i.e. for $S$ consecutive OFDM symbols.}. 
The signal received at the $k$-th UE, before the receive beamforming is applied, after some algebraic manipulations, can be then written as
\begin{equation}
	\begin{array}{llll}
		\br_{k,s}(t) =& \ds \sum _{m=1}^M \ds \int \bH_{k,m}^{(s)}(\tau) \bx_{m,s}(t-\tau) d \tau  + \bz_{k,s}(t)= \\ &
		\ds \sum _{m=1}^M \sum_{\ell=0}^{L_{k,m}}  \sum_{i=1}^{n_{\rm AP}} \alpha_{k,m}^{(s)}(\ell) g^{(AP)}_{k,m,\ell,s,i} \times \\ & \;  x_{m,s,i}(t-\tau_{k,m,\ell}) \ba_{\rm UE}(\varphi_{k,m,\ell}) + \bz_{k,s}(t) \; ,
	\end{array}
	\label{eq:received_ks}
\end{equation}
with $g^{(AP)}_{k,m,\ell,s,i}=\ba_{\rm AP}^H(\theta_{k,m,\ell})\bu_{m,s,i}$ and $\bz_{k,s}(t)$ an $N_{MS}$-dimensional vector waveform representing the AWGN contribution at the $k$-th UE receiver in the $s$-th beacon interval.

Since the $k$-th UE is equipped with $n_{\rm UE}$ RF chains, it can apply $n_{\rm UE}$ different receive beamforming vectors to the received signal in Eq. \eqref{eq:received_ks}. Denoting by $\bv_{k,s,j}$ the $j$-th beamformer (with $j=1, \ldots, n_{\rm UE}$) used by the $k$-th UE in the $s$-th beacon slot, the following set of observables is available at the $k$-th UE after beamforming:
\begin{equation}
	\begin{array}{llll}
		y_{k,s,j}(t)= &\ds \frac{1}{\sqrt{n_{\rm UE}}} \bv^H_{k,s,j}\br_{k,s}(t) = \\ &
		\ds \sum _{m=1}^M \sum_{\ell=0}^{L_{k,m}}\sum_{i=1}^{n_{\rm AP}} \ds \frac{1}{\sqrt{n_{\rm UE}}}\alpha_{k,m}^{(s)}(\ell)
		{g^{(AP)}_{k,m,\ell,s,i}} g^{(UE)}_{k, \ell, s, j} \times \\ & \;
		x_{m,s,i}(t-\tau_{k,m,\ell}) + z_{k,s,j}(t) \; ,
	\end{array}
	\label{eq:received_ksj}
\end{equation}
for $j=1, \ldots, n_{\rm UE}$, with $g^{(UE)}_{k, \ell, s, j}= \bv^H_{k,s,j}\ba_{\rm UE}(\varphi_{k,m,\ell})$ and $z_{k,s,j}(t)= \frac{1}{\sqrt{n_{\rm UE}}}\bv^H_{k,s,j} \bz_{k,s}(t)$.
In \eqref{eq:received_ksj}, the factor $\frac{1}{\sqrt{n_{\rm UE}}}$ accounts for the effect of the signal splitter that divides the useful received power into $n_{\rm UE}$ equal-power parts.
The waveforms $y_{k,s,j}(t)$, for all $j$, undergo the usual OFDM receiver processing, i.e., every OFDM symbol in $y_{k,s,j}(t)$ is converted into a vector with $N_C$ entries. Focusing on the generic $p$-th OFDM symbol, and letting $s(p)=\lfloor p/S \rfloor$ denote the  beacon slot index associated with the $p$-th OFDM symbol, the A/D conversion leads to the scalar entries 
say $Y_{k,p,j}(0), \ldots, Y_{k,p,j}(N_C-1)$.
In particular, it is easy to see that the $q$-th entry of such vector is expressed as
\begin{equation}
	\begin{array}{llll}
		Y_{k,p,j,i}(q)=&  \ds \frac{1}{\sqrt{n_{\rm UE}}} \ds  \sum _{m=1}^M \bv^H_{k,s(p),j}  
		\mathcal{H}_{k,m}^{(s)}(q)\times \\ & \;  X_{m,p,i}(q) \bu_{m,s(p),i} + Z_{k,p,j,i}(q) \; ,
	\end{array}
	\label{eq:Yskj}
\end{equation}
where $X_{m,p,i}(q)$ is the $q$-th data symbol transmitted in the $p$-th OFDM slot on the $i$-th transmit RF chain, $Z_{k,p,j,i}(q)$ contains the AWGN contribution and $\mathcal{H}_{k,m}^{(s)}(q)$ is the matrix-valued Fourier transform of the channel impulse response $\bH_{k,m}^{(s)}(\tau)$ computed at the frequency $q/(t_0)$, i.e.,

\begin{equation*}
	\mathcal{H}_{k,m}^{(s)}(q)\!\!=\!\!\sum_{\ell=0}^{L_{k,m}} \!\! \alpha_{k,m}^{(s)}(\ell) \ba_{\rm UE}(\varphi_{k,m,\ell}) \ba_{\rm AP}^H(\theta_{k,m,\ell})
	e^{-\i 2\pi \frac{q}{t_0}\tau_{k,m,\ell}}\, ,
\end{equation*}
where $\i$ is the imaginary unit.

\subsection{Signal discretization at the UE}
The AoA and AoDs,  $\varphi_{k,m,\ell}$ and $\theta_{k,m,\ell}$ in Eq. \eqref{eq:channelmodel_km}, respectively, take continuous values, but in the BA procedure we use the approximate finite-dimensional (discrete) beamspace representation following the approaches in  \cite{Heath_SP_mmwave_2016,Caire_scalable_robust_BA_TCOM2018}. We consider the discrete set of AoDs and AoAs
\begin{equation}
	\begin{array}{llll}
		&\Theta=\left\lbrace \widehat{\theta} : \ds \frac{1 + \sin( \widehat{\theta} ) }{2} = \ds \frac{u-1}{N_{ \rm AP}}, \; u=1,\ldots, N_{\rm AP} \right \rbrace \, , \\
		&\Phi=\left\lbrace \widehat{\varphi} : \ds \frac{1 + \sin( \widehat{\varphi} ) }{2} = \ds \frac{u'-1}{N_{\rm UE}}, \; u'=1,\ldots, N_{\rm UE} \right \rbrace
	\end{array}
\end{equation}
and use the corresponding array responses $\mathcal{A}= \left \lbrace  \ba ( \widehat{\theta}) : \widehat{\theta} \in \Theta \right \rbrace $ and $\mathcal{B}= \left \lbrace  \bb ( \widehat{\varphi}) : \widehat{\varphi} \in \Phi \right \rbrace $ as a discrete dictionary to represent the channel response. For the ULAs considered in this approach the dictionaries $\mathcal{A}$ and $\mathcal{B}$, after suitable normalization, yield orthonormal bases corresponding to the columns of the unitary discrete Fourier transform (DFT) matrices $\bW_{N_{\rm AP}}$ and $\bW_{N_{\rm UE}}$ defined as 
\begin{equation}
	\begin{array}{lll}
		\left[\bW_{N}\right]_{p,p'}=\ds \frac{1}{\sqrt{N}}e^{ \i 2 \pi (p-1)\left(\frac{p'-1}{N} -\frac{1}{2}\right)} & p, p'=1, \ldots, N,
	\end{array}
\end{equation}
with $N \in \left\lbrace N_{\rm AP},N_{\rm UE}\right \rbrace$.
The columns of $\bW_{N_{\rm AP}}$ ($\bW_{N_{\rm UE}}$) represent an orthonormal basis for 
$\mathcal{C}^{N_{\rm AP}}$ ($\mathcal{C}^{N_{\rm UE}}$); as a consequence, we can define
\begin{equation}
	\begin{array}{lll}
		\mathbb{v}_{k,s,j}= \ds \bW^H_{N_{\rm UE}} \bv_{k,s,j}, \quad 
		\mathbb{u}_{m,s,i}= \ds \bW^H_{N_{\rm AP}} \bu_{m,s,i}, \\
		\mathbb{H}^{(s)}_{k,m}(q)= \ds \bW^H_{N_{\rm UE}} \mathcal{H}^{(s)}_{k,m}(q) \bW_{N_{\rm AP}}
	\end{array}
\end{equation}
and \eqref{eq:Yskj} can be shown to be written as
\begin{equation}
	\begin{array}{llll}
		Y_{k,p,j,i}(q)=& \ds \frac{1}{\sqrt{n_{\rm UE}}} \ds  \sum _{m=1}^M \mathbb{v}^H_{k,s(p),j}
		\mathbb{H}^{(s)}_{k,m}(q) \mathbb{u}_{m,s(p),i} \times \\ & \; 
		X_{m,s(p),i}(q)  + Z_{k,p,j,i}(q) \; , \label{eq:Yskj2}
	\end{array}
\end{equation}
The beamforming vector $\bu_{m,s,i}$ to be used at the $m$-th AP in the $s$-th beacon slot and on the $i$-th RF chain is defined by the data pattern associated at the AP. In general, it will be an all-zero vector with $\nu_{\rm AP}$ entries equal to one; the positions of the ones are pseudo-random and defined by the data-pattern. Similarly, the beamformer $\bv_{k,s,j}$ to be used at the $k$-th UE in the $s$-th beacon slot and on the $j$-th RF chain will be an all-zero vector with $\nu_{\rm UE}$ ones placed in pseudo-random positions. 

The use of the DFT matrices and of the beamspace representation is useful in order to obtain a discrete BA problem. Indeed, at the design stage, we assume that the AoAs and AoDs of the strongest multipath components coincide with the discrete angles represented by the columns of the DFT matrices. In practice, of course, this cannot be guaranteed and there will be some approximation error. The error, however, is lower and lower as the number of antenna elements increases.

During the BA phase, each AP transmits a positive constant value, i.e. $X_{m,s,i}(q)=\sqrt{\beta}$, on its assigned subcarriers for $T$ consecutive beacon slots. Each UE can rely on the knowledge of the data patterns ${\cal D}_1, \ldots, {\cal D}_D$, and, based on them, has to determine the AoA and AoD of the strongest multipath components to be used for data communication. Notice that no information on the APs location or on the network topology is needed at the UE. Similarly, the proposed BA procedure is non-coherent, in the sense that no specific pilot sequences are to be transmitted by the APs.

First of all, we notice that since each data pattern adopts a disjoint set of subcarriers, the UE can operate on $D$ different sets of observables. More precisely, letting ${\cal A}_d \in \{ 1, 2, \ldots, M\}$ denote the set of APs that has been assigned the $d$-th data pattern, the observables in Eq. \eqref{eq:Yskj2} can be regrouped in $D$ disjoint sets to be separately processed. 
The $d$-th set of observables available at the $k$-th UE, that we denote by ${\cal O}_k^{(d)}$, is thus expressed as
\begin{equation*}
\begin{array}{llll}
{\cal O}_k^{(d)} &=\left\{\overline{Y}^{(d)}_{k,p,j,i}(q), \; q \in {\cal L}_{d,s,i}, \;  s=1, \ldots, T,  \right. \\ & p=1, \ldots, ST,
\left.  \; i=1, \ldots, n_{\rm AP}, \; j=1, \ldots, n_{\rm UE} \rule{0mm}{7mm} \right\},
\label{eq:finaldata_d}
\end{array}
\end{equation*}
with
\begin{equation}
\begin{array}{llll}
\overline{Y}^{(d)}_{k,p,j,i}(q)=& \ds \frac{\sqrt{\beta}}{\sqrt{n_{\rm UE}}} \ds  \sum _{m \in \mathcal{A}_d}  \mathbb{v}^H_{k,s(p),j}
\mathbb{H}^{(s)}_{k,m}(q) \mathbb{u}_{m,s(p),i} \\ &  + \overline{Z}^{(d)}_{k,p,j}(q)\, .
\end{array}
\end{equation}
Based on the above data, the following averaged quadratic observable is built:
\begin{equation}
c_{k,s,j,i}^{(d)}=\ds\frac{1}{Q S}\ds \sum_{p=(s-1)S+1}^{sS}\sum_{q\in {\cal L}_{d,s,i}}\left|\overline{Y}^{(d)}_{k,p,j,i}(q)\right|^2 \; .
\label{eq:averaged}
\end{equation}
We now propose two different algorithms to extract the information on the AoA and AoD of the strongest path from the closest AP using the $d$-th data pattern.
\subsubsection{Stacked collection of observables (SCO)}
This algorithm is inspired by the one in \cite{Caire_scalable_robust_BA_TCOM2018} for a single-AP system. First of all, the measurements are collected for all the values of $i$, $j$ and $s$ and grouped into the following vector:
\begin{equation}
\mathbf{c}_k^{(d)}\!=\!\!\left[c_{k,1,1,1}^{(d)}, \ldots, c_{k,1,n_{\rm UE},n_{\rm AP}}^{(d)}, c_{k,2,1,1}^{(d)}, \ldots, 
c_{k,T,n_{\rm UE},n_{\rm AP}}^{(d)}\right]^T  .  
\end{equation}
Next, let 
\begin{equation}
\mathbb{b}_{k,s,j,i}^{(d)}=\ds \frac{\mathbb{d}_{d,s,i} \otimes \mathbb{v}^H_{k,s,j}}{\| \mathbb{d}_{d,s,i} \| \| \mathbb{v}^H_{k,s,j}\|} \; ,
\end{equation}
and form the $(n_{\rm AP}n_{\rm UE}T \times N_{\rm AP}N_{\rm UE})$-dimensional matrix 
\begin{equation}
	\begin{array}{ll}
\mathbf{B}_k^{(d)} = & \left[\mathbb{b}_{k,1,1,1}^{(d)}, \ldots, \mathbb{b}_{k,1,n_{\rm UE},n_{\rm AP}}^{(d)}, \right. \\ & \left.  \mathbb{b}_{k,2,1,1}^{(d)}, \ldots, 
\mathbb{b}_{k,T,n_{\rm UE},n_{\rm AP}}^{(d)}\right]^T \; .
\end{array}\end{equation}
Based on the above notation, the following optimization problem can be considered:
\begin{equation}
\bm{\xi}_{k}^{(d)}= \mbox{arg} \, \min_{\mathbf{x}} \left\| \mathbf{B}_k^{(d)} \mathbf{x} + \sigma^2 \mathbf{1}_{n_{\rm AP}n_{\rm UE}T\times 1} - \mathbf{c}_k^{(d)}\right\|^2 \; .
\label{eq:problem_kd}
\end{equation}
The solution $\bm{\xi}_{k}^{(d)}$ to Problem \eqref{eq:problem_kd}  is a $(N_{\rm AP}N_{\rm UE})$-dimensional vector that can be arranged in a $(N_{\rm AP}\times N_{\rm UE})$-dimensional matrix, $\bm{\Xi}_{k}^{(d)}$ say, where each entry can be associated to a pair (AoD, AoA) associated to a possible propagation path coming from the APs using the $d$-th data pattern. Each entry of $\bm{\Xi}_{k}^{(d)}$ contains, for each possible pair (AoD, AoA), an estimate of the channel power; it thus follows that the largest entry of  $\bm{\Xi}_{k}^{(d)}$ is an indicator of the dominant path between the $k$-th UE and the APs using the $d$-th data pattern. If the data patterns have been carefully assigned to the APs, it is reasonably expected that each UE will receive a dominant contribution from only one of these APs, while the contributions from the remaining APs can be assimilated to background noise.
The $k$-th UE detects the $N_D \leq D$ strongest (AoA, AoD) pairs which can be thus used for data communication. This procedure is independently implemented by each UE, i.e., $k=1,\ldots,K$.

\subsubsection{Matrix-valued collection of observables (MCO)} 
The measurements can be collected for all the values of $i$, $j$ and $s$ and grouped in the $(N_{\rm UE}\times N_{\rm AP})$-dimensional matrix $\mathbf{C}_k^{(d)}$ whose generic entry is
\begin{equation}
\left(\mathbf{C}_k^{(d)}\right)_{(\ell, \ell')}=\ds \sum_{s=1}^T I_{\ell, \ell'}c_{k,i,j,s}^{(d)}
\label{C_matrix}
\end{equation}
where $I_{\ell, \ell'}$ is one if the $\ell'$-th transmit direction at the AP and the $\ell$-th receive direction at the $k$-th UE are involved in the measurement of $c_{k,i,j,s}^{(d)}$. Given the matrix $\mathbf{C}_k^{(d)}$ obtained in Eq. \eqref{C_matrix},  the positions of its largest entry is an indicator of the dominant path between the $k$-th UE and the AP using the $d$-th data pattern. Note that in the MCO algorithm, no optimization problem is to be solved to obtain the largest entry of matrix $\mathbf{C}_k^{(d)}$; conversely,  in the SCO algorithm, problem \eqref{eq:problem_kd} must be solved in order to find matrix $\bm{\Xi}_{k}^{(d)}$. 
The complexity of the MCO procedure is thus considerably lower than that of the SCO procedure.

\begin{figure}
	\begin{center}
		\includegraphics[scale=0.5]{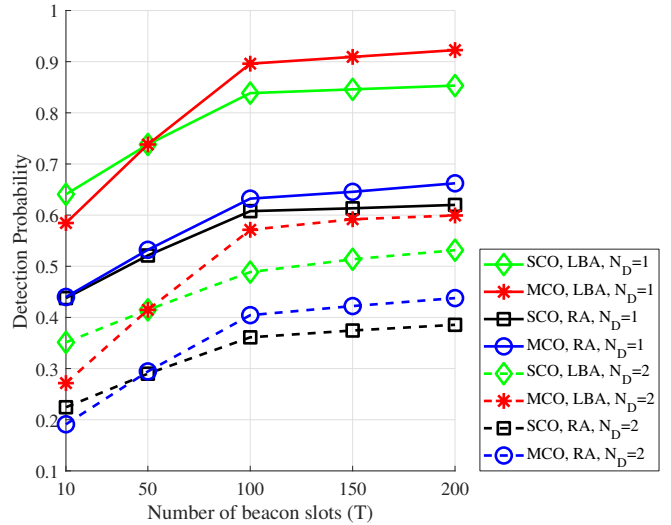}
	\end{center}
	\caption{Detection probability of the BA proposed schemes MCO ans SCO versus the number of beacon slots $T$ for two values of $N_D$. Parameters: $M=50, K=15, N_{\rm AP}=32$, $N_{\rm UE}=16$, $D=8$, $n_{\rm AP}=8$, $n_{\rm UE}=4$, $\nu_{\rm AP}=8$, and $\nu_{\rm UE}=4$.}
	\label{Fig_Detection_prob_MCO_SCO}
\end{figure}

\begin{figure*}
	\begin{center}
		\includegraphics[scale=0.5]{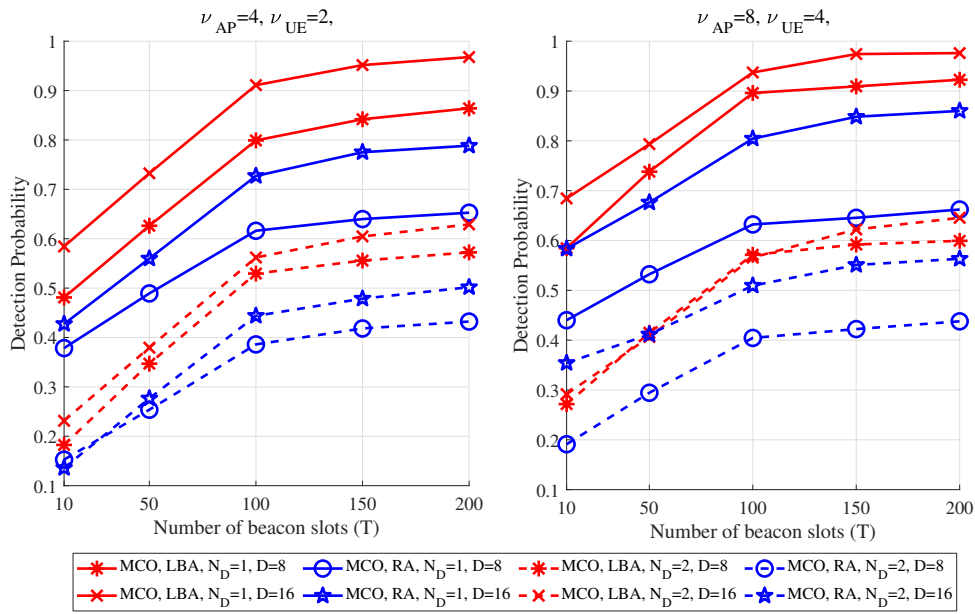}
	\end{center}
	\caption{Detection probability of the MCO BA proposed scheme versus the number of beacon slots $T$ for different values of $N_D$, $\nu_{\rm AP}$, $\nu_{\rm UE}$ and $D$. Parameters: $M=50, K=15, N_{\rm AP}=32$, $N_{\rm UE}=16$, $n_{\rm AP}=8$, and $n_{\rm UE}=4$.}
	\label{Fig_Detection_prob_MCO_D}
\end{figure*}

\section{Numerical Results}
\subsection{Simulation setup}
In our simulation setup, we assume a communication bandwidth $W = 500$ MHz centered over the carrier frequency $f_0=28$ GHz. The OFDM subcarrier spacing is 480 kHz and assuming that the length of the cyclic prefix is 7\% of the OFDM symbol duration, i.e., $\tau_{\rm CP}\Delta_f=0.07$, we obtain $t_0=2.23 \mu$s and $N_C=1024$ subcarriers. A beacon slot contains $S=14$ OFDM symbols and the random access slot, used to share the results of the BA procedure, also contains 14 OFDM symbols\cite{Caire_scalable_robust_BA_TCOM2018}. The antenna height at the AP is $10$ m and at the UE is $1.65$ m. The additive thermal noise is assumed to have a power spectral density of $-174$ dBm/Hz, while the front-end receiver at the APs and at the MSs is assumed to have a noise figure of $9$ dB. We consider a square area of 400m $\times$ 400 m, the number of UE is $K=15$ and the number of APs is $M=50$; the APs and UEs are equipped with ULAs of $N_{\rm AP}=32$ and $N_{\rm UE}=16$ antennas, respectively, the number of RF chains at the APs and UEs are $n_{\rm AP}=8$ and $n_{\rm UE}=4$, respectively. We assume a number of total scatterers, $N_{\rm s}=300$ say, common to all the APs and UEs and uniformly distributed in the simulation area. In order to model the signal blockage, we assume that the communication between the $m$-th AP and the $k$-th UE takes place via the $n$-th scatterer, i.e., the $n$-th scatterer is one of the effective $L_{k,m}$ contributing in the channel in Eq. \eqref{eq:channelmodel_km}, if the rays between the $m$-th AP and the $n$-th scatterer and the $k$-th UE and the $n$-th scatterer \emph{simultaneously} exist. We assume that a link exists between two entities, in our case one AP/UE and one scatterer, if they are in LoS, with a LoS probability given by \cite{ghosh20165g_WP,5G3PPPlikemodel}. For the channel model in Eq. \eqref{eq:channelmodel_km}, the variance of the complex gain associated to the $\ell$-th path between the $k$-th user and the $m$-th AP, $\gamma_{k,m}(\ell)$ is obtained as reported in  \cite{ghosh20165g_WP}. The total power transmitted by the APs over all the subcarriers during the BA phase is denoted as $P_{\rm BA}$, and consequently $\beta=P_{\rm BA}/N_C$. In the following, we assume $P_{\rm BA}=7$ dBW.

\subsection{Detection Probability}

The considered performance measure is the probability of correct detection at the UE of the AoA and AoD of the $N_D$ strongest paths (one for each AP), i.e. the probability that a UE detects the correct AoA and AoD for the strongest path from the $N_D$ best APs. 
Fig. \ref{Fig_Detection_prob_MCO_SCO} shows such detection probability versus the number of used beacon slots $T$. In this figure, there are $D=8$ different data patterns, while the number of active fingers in the beamformers is $\nu_{\rm AP}=8$ and $\nu_{\rm AP}=4$. In order to show the merits of the proposed LBA data pattern assignment procedure, we also report the performance corresponding to a random assignment (RA) of the data patterns to the APs. We can see that the MCO, albeit being simpler, achieves much better performance than the SCO. For this reason, in Figs. \ref{Fig_Detection_prob_MCO_D} we only report the MCO algorithm performance, assuming different values of $\nu_{\rm AP}$ and $\nu_{\rm UE}$ and with $D=8$ and $D=16$. Inspecting the figure, we can see that the increase in the parameter $D$ improves the detection capability of the system; however $D$ cannot be increased too much since this corresponds to a smaller value of $Q$, the number of carriers assigned to each AP RF chain. Larger values for the parameters $\nu_{\rm AP}$ and $\nu_{\rm UE}$ also bring some performance improvement  in the case of low values of $T$. Indeed, increasing $T$, we have a longer time to test the different antennas in the system and consequently it appears that is better to use few active fingers with higher transmit power than more fingers with lower transmit power.
Overall, the results show that the proposed procedures are effective and permit realizing BA in multi-AP multi-UE environments with satisfactory performance.

\section{Conclusions}
This paper has considered the problem of performing BA in a CF-mMIMO network operating at mmWave. The proposed BA procedure amounts to a protocol involving the CPU, the UEs, the APs and a macro-BS managing a control channel at sub-6 GHz frequency. It enables simultaneous BA of each UE with the strongest beams coming from a pre-defined number of strongest APs. 
A procedure in order to assign the data patterns across the APs has also been proposed, and two different algorithms, to be run at the UE, have been proposed. Of these, the MCO has been shown to achieve better performance and with less complexity than the other proposed algorithm, the SCO one, that was inspired by \cite{Caire_scalable_robust_BA_TCOM2018}.
Numerical results have confirmed the effectiveness of the proposed approach, that has shown that BA can be performed in a shared frequency band with a simultaneous operation of several APs and several UEs. 

\section*{Acknowledgement}
This work was supported by HiSilicon through cooperation agreement YBN2018115022.

\bibliography{references}
\bibliographystyle{IEEEtran}

\end{document}